\documentclass[sigconf]{acmart}
\AtBeginDocument{%
  }

\usepackage{algorithm}
\usepackage{algorithmic}

\usepackage{amsmath}
\usepackage{wasysym}
\usepackage{subfig}
\usepackage{multirow}
\usepackage{xspace}
\usepackage{threeparttable}
\usepackage{enumitem}
\usepackage{amsfonts}
\let\oldhat\hat
\renewcommand{\vec}[1]{\mathbf{#1}}    
\renewcommand{\hat}[1]{\oldhat{\mathbf{#1}}}

\newcommand{\ie}{\emph{i.e., }}
\newcommand{\eg}{\emph{e.g., }}

\newcommand{\etc}{\emph{etc}}
\newcommand{\wrt}{\emph{w.r.t. }}

\newcommand{\aka}{\emph{aka. }}

\copyrightyear{2024}
\acmYear{2024}
\setcopyright{acmlicensed}
\acmConference[WWW '24 Companion] {Companion Proceedings of the ACM Web Conference 2024}{May 13--17, 2024}{Singapore, Singapore.}
\acmBooktitle{Companion Proceedings of the ACM Web Conference 2024 (WWW '24 Companion), May 13--17, 2024, Singapore, Singapore}
\acmISBN{979-8-4007-0172-6/24/05}
\acmDOI{10.1145/3589335.3648296} 
\begin{document}

\title{Knowledge Enhanced Multi-intent Transformer Network for Recommendation}
\author{Ding Zou}
\authornote{Work done during internship at Taotian Group.}
\affiliation{%
\department[0]{CCIIP Lab}
\department[1]{School of Computer Science and Technology}
  \institution{Huazhong University of Science and Technology}
  \institution{Joint Laboratory of HUST and Pingan Property \& Casualty Research (HPL)}
  \city{Wuhan}
  \country{China}}
\affiliation{
  \institution{Taotian Group}
  \city{Hangzhou}
  \country{China}}
\email{m202173662@hust.edu.cn}

\author{Wei Wei}
\authornote{Corresponding author.}
\affiliation{
\department[0]{CCIIP Lab}
\department[1]{School of Computer Science and Technology}
\institution{Huazhong University of Science and Technology}
\institution{Joint Laboratory of HUST and Pingan Property \& Casualty Research (HPL)}
  \city{Wuhan}
  \country{China}}
\email{weiw@hust.edu.cn}


\author{Feida Zhu}
\affiliation{%
  \institution{Singapore Management University}
  \city{Singapore}
  \country{Singapore}}
\email{fdzhu@smu.edu.sg}

\author{Chuanyu Xu}
\affiliation{%
 \institution{Taotian Group}
 \city{Hangzhou}
 \country{China}}
\email{tracy.xcy@taobao.com}

\author{Tao Zhang}
\affiliation{%
  \institution{Taotian Group}
  \city{Hangzhou}
  \country{China}}
\email{guyan.zt@taobao.com}

\author{Chengfu Huo}
\affiliation{%
  \institution{Taotian Group}
  \city{Hangzhou}
  \country{China}}
\email{chengfu.huocf@taobao.com}

\renewcommand{\shortauthors}{Ding Zou et al.}

\begin{abstract}
Incorporating Knowledge Graphs (KGs) into Recommendation has attracted growing attention in industry, due to the great potential of KG in providing abundant supplementary information and interpretability for the underlying models.
However, simply integrating KG into recommendation usually brings in negative feedback in industry, mainly due to the ignorance of the following two factors:
i) users' multiple intents, which involve diverse nodes in KG. For example, in e-commerce scenarios, users may exhibit preferences for specific styles, brands, or colors.
ii) knowledge noise, which is a prevalent issue in Knowledge Enhanced Recommendation (KGR) and even more severe in industry scenarios. The irrelevant knowledge properties of items may result in inferior model performance compared to approaches that do not incorporate knowledge.
To tackle these challenges, we propose a novel approach named \underline{K}nowled\underline{g}e Enhanced Multi-intent \underline{T}ransformer \underline{N}etwork for Recommendation (KGTN), which comprises two primary modules: Global Intents Modeling with Graph Transformer, and Knowledge Contrastive Denoising under Intents.
Specifically, Global Intents with Graph Transformer focuses on capturing learnable user intents, by incorporating global signals from user-item-relation-entity interactions with a well-designed graph transformer, and meanwhile learning intent-aware user/item representations.  
On the other hand, Knowledge Contrastive Denoising under Intents is dedicated to learning precise and robust representations. It leverages the intent-aware user/item representations to sample relevant knowledge, and subsequently proposes a local-global contrastive mechanism to enhance noise-irrelevant representation learning.
Extensive experiments conducted on three benchmark datasets show the superior performance of our proposed method over the state-of-the-arts. 
And online A/B testing results on Alibaba large-scale industrial recommendation platform also indicate the real-scenario effectiveness of KGTN.
The implementations are available at: https://github.com/CCIIPLab/KGTN.
\end{abstract}


\maketitle

\section{Introduction}

\begin{figure}[th] 
    \centering  
    \subfloat[Intent Case]
    {
        \begin{minipage}[t]{0.56\linewidth}
            \centering      
            \includegraphics[width=\textwidth]{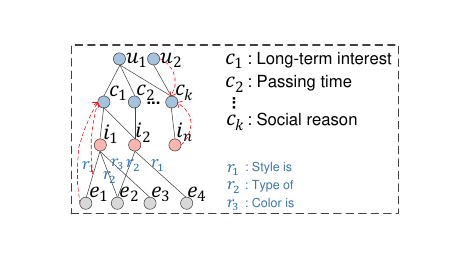}
        \end{minipage}
    }
    \subfloat[Comparison]
    {
        \begin{minipage}[t]{0.4\linewidth}
            \centering     
            \includegraphics[width=\textwidth]{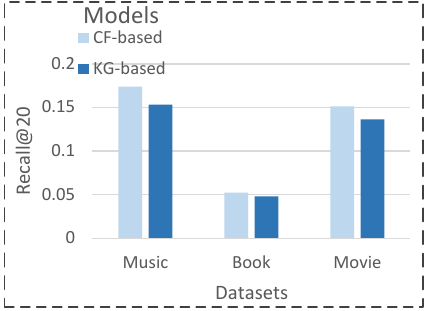}   
        \end{minipage}
    }
    \caption{(a) A simple case for illustrating multiple user intents with global information; (b)Performance comparison.} 
    \label{fig::toy} 
\end{figure}

Knowledge graphs (KGs) have emerged as a promising approach to enhance the accuracy and interpretability of recommender systems in both academic and industry scenarios. By incorporating entities and relations, KGs provide a rich source of information for user/item representation learning, which not only captures the diverse relationships among items (such as the same item brand), but also allows for the interpretation of user preferences (such as attributing a user's selection of a clothing to its fashionable style). 

In an effort to effectively integrate the item-side KG information into recommendation, considerable research efforts have been devoted to Knowledge Enhanced Recommendation (\aka KGR).
Early studies \cite{zhang2016collaborative, huang2018improving,wang2018dkn} directly integrate knowledge graph embeddings with items to enhance their representations. 
Some subsequent studies \cite{hu2018leveraging, shi2018heterogeneous, wang2019explainable} enrich the interactions via meta-paths that capture relevant connectivities between users and items with KG. They either select prominent paths over KG \cite{sun2018recurrent}, or represent the interactions with multi-hop paths from users to items \cite{hu2018leveraging, wang2019explainable}.
Nevertheless, most of them heavily rely on manually designed meta-paths, which makes it hard to optimize in reality. 
As a result, later methods have embraced Graph Neural Networks (GNNs) \cite{wang2021learning, wang2019kgat} to automatically aggregate high-order information over KG, which iteratively integrate multi-hop neighbors into representations and have demonstrated promising performance for recommendation.
Most recently, there have been efforts to incorporate Contrastive Learning (CL) into KGR for addressing noisy knowledge and long-tail problems \cite{yang2022knowledge, zou2022multi, wang2024exploring} via contrasting the user-item (collaborative part) and item-entity (knowledge part) graphs.


However, current KGR methods usually bring poor performance in large-scale industry scenarios, due to their commonly overlooking two crucial factors:
1) Users' multiple intents underlying interaction behavior. For instance, as depicted in Figure \ref{fig::toy}(a), users may have diverse intentions when shopping in Alibaba E-commerce platform, such as long-term interest, passing time, or social reason, \etc.
2) Redundant Knowledge information.
In the context of user intents, some knowledge facts in the KG may be irrelevant noise \cite{chen2022attentive}, which can potentially disrupt the learning process of user/item representations. As shown in Figure \ref{fig::toy}(b), incorporating KGs may result in a worse model performance than the models without KG utilization (the details of comparison could refer to Section~\ref{compare_exp} ).

But still, it's not trivial to model user intents in KGR, since user intents may be composed of multiple heterogeneous information, including items, relations, and entities. Previous multi-intent modeling methods usually define the intents as a linear combination of either interacted items \cite{wang2020disentangled} or entire relation sets \cite{wang2021learning}, then update the intent representations through local aggregation in the user-intent-item heterogeneous graph. Nevertheless, such a multi-intent learning paradigm may not fully meet the requirements for KGR, as it neglects the global information in intent defining and learning. To illustrate this, we present an example in Figure \ref{fig::toy}(a). In this example, user $u_1$ may purchase the item $i_1$ for the intent $c_1$ of long-term interest, resulting in a focus on clothing style (\eg whether it is fashionable), which means intent $c_1$ is associated with KG relation $r_1$ and entity $e_1$; while $u_1$ may buy the item $i_n$ for the intent $c_k$ of social reason (such as friend $u_2$ recommend), which means intent $c_k$ is associated with user $u_2$ and item $i_k$.


In this paper, we focus on modeling user intents behind interaction behaviors with global collaborative (user-item) and knowledge (item-relation-entity) information, and exploiting these modeled intents to guide knowledge sampling, facilitating fine-grained and accurate user/item representation learning. We propose a novel model, KGTN, which comprises two essential components for solving the foregoing limitations:
i) Global Intents Modeling with Graph Transformer. We predefine $K$ intent representations for user/item, then learn these intents with global information from collaborative and knowledge graphs. Specifically, it first merges knowledge information into items, then propose a novel graph transformer in the user-item graph to learn global intents and generate intent-aware user/item representations.
ii) Knowledge Contrastive Denoising under Intents. KGTN first exploits the intent-aware user/item representations to guide the knowledge sampling, effectively pruning the irrelevant knowledge. Then a novel local-global contrastive mechanism is proposed here to denoise the user/item representations.
Empirically, KGTN outperforms the state-of-the-art models on three benchmark datasets in offline testing, and achieves significant improvements in online A/B testing.

\textbf{Our contributions} of this work can be summarized as follows:
\begin{itemize}[leftmargin=*]
    \item \textbf{General Aspects:} We emphasize the importance of intent modeling with global information, which plays a crucial role in fine-grained representation learning and knowledge denoising.
    \item \textbf{Novel Methodologies:} We propose a novel model KGTN, which models user intents from global signals with a novel graph transformer; and denoises item representations with i) knowledge denoising under intents, and ii) local-global graph contrastive learning.
    \item \textbf{Multifaceted Experiments:} We conduct extensive offline experiments on three benchmark datasets and online A/B testing on Alibaba recommendation platform. The results demonstrate the advantages of our KGTN in better representation learning.
\end{itemize}

\section{Problem Formulation}

In this section, we begin by formulating the structural data of CF (user-item interactions) and KG (item-relation-entity knowledge) in KGR, then present the problem statement.

\noindent\textbf{Interaction Data}. 
In a typical recommendation scenario, let $\mathcal{U}=\left\{u_{1}, u_{2}, \ldots, u_{M}\right\}$ be a set of $M$ users and $\mathcal{V}=\left\{v_{1}, v_{2}, \ldots, v_{N}\right\}$ a set of $N$ items.
Let $\mathbf{Y} \in \mathbf{R}^{M \times N}$ be the user-item interaction matrix, where $y_{u v}=1$ indicates that user $u$ engaged with item $v$, such as behaviors like clicking or purchasing; otherwise $y_{u v}=0$.

\noindent\textbf{Knowledge Graph}. 
A KG stores luxuriant real-world facts associated with items, encompassing item attributes or external commonsense knowledge, in the form of a heterogeneous graph \cite{shi2018heterogeneous}. 
Let $\mathcal{G}=\{(h, r, t) \mid h, t \in \mathcal{E}, r \in \mathcal{R}\}$ be the KG, where $h$, $r$, $t$ represent the head, relation, tail of a knowledge triple, respectively; $\mathcal{E}$ and $\mathcal{R}$ denote the sets of entities and relations in $\mathcal{G}$. 
In many recommendation scenarios, an item $v \in \mathcal{V}$ corresponds to one entity $e \in \mathcal{E}$. We hence establish a set of item-entity alignments $\mathcal{A} =\{(v, e)|v \in \mathcal{V}, e \in \mathcal{E}\}$, where $\left(v, e\right)$ indicates that item $v$ can be aligned with an entity $e$ in KG.
With the alignments between items and KG entities, KG is able to profile items and offer complementary information to the interaction data.

\noindent\textbf{Problem Statement}. 
Given the user-item interaction matrix $\mathbf{Y}$ and the KG $\mathcal{G}$, KGR aims to learn a function that can predict how likely a user would adopt an item.

\section{Methodology}

\begin{figure*}[th]
  \centering
  \includegraphics[width=\textwidth]{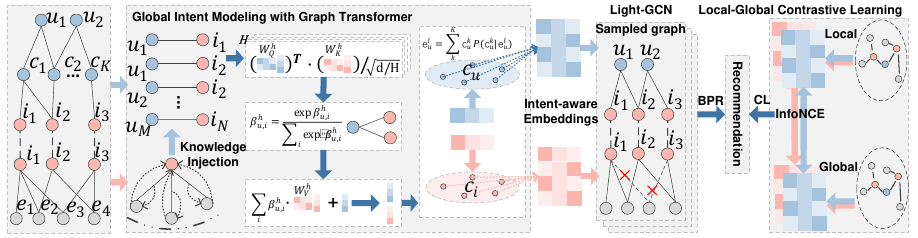}
  \caption{Overall framework illustration of the proposed KGTN model. Best viewed in color.}
  \label{fig:model}
\end{figure*}

We now present the proposed \underline{K}nowled\underline{g}e Enhanced Multi-intent \underline{T}ransformer \underline{N}etwork for Recommendation (KGTN). KGTN aims at modeling user intents with global information and exploiting user intents to denoise KG for accurate and robust user/item representation learning.
Figure \ref{fig:model} displays the framework of KGTN, which mainly consists of two key components: 
1) Global Intent Modeling with graph transformer.
Initially, KGTN defines a set of $K$ learnable global intents for users and items. It then models these intents and learns intent-aware user/item representations, via integrating global signals with a graph transformer in the user-item graph, where knowledge information has been encoded into items. 
2) Knowledge Contrastive Denoising under intents. 
It first exploits the learned intent-aware user/item representations to sample intent-relevant knowledge, then designs a contrastive self-supervised task between the local aggregation and global aggregation features within the sampled graph to facilitate robust representation learning. 

\subsection{Global Intents Modeling with Graph Transformer}

\subsubsection{Intent Initialization with Global signals}

When interacting with items, users often have diverse intents, such as preferences for specific clothing brands and styles, friends recommending, or passing time with randomly clicking~\cite{wang2021learning, ren2023disentangled}. To capture these diverse intents, we assume $K$ different intents $c_u$ and $c_v$ from the user and item sides, respectively, where the intents on the item side can also be understood as the theme or context of the item, for example, a user who intends to purchase a fashionable dress may like clothes of ``young'' topic. 
Our predictive objective of user-item preference can be presented as follows:
\begin{align}\label{eq4}
    \int_{c_{u}}\ \int_{c_{v}}\ P(y, c_{u}, c_{v}|u, v)\, dc_{v}\, dc_{u} = \sum_{k}^{K} P(y, c_{u}^{k}, c_{v}^{k}|u, v).
\end{align}

Specifically, we define $K$ global intent prototypes $\{\textbf{c}_{u}^{k}\in\mathbb{R}^{d}\}_{k=1}^K$ and $\{\textbf{c}_{v}^{k}\in\mathbb{R}^{d}\}_{k=1}^K$ for user and item, respectively. With these predefined intent prototypes, we then are supposed to integrate them into user/item representations, and update them with related global signals.

\subsubsection{Intent Modeling with graph transformer}
Towards accurately modeling user intents with global information and learning intent-aware user/item representations, we perform an intent-aware information propagation with these learnable intents. Specifically, intent-aware user/item embeddings are acquired by an attentive sum of the intent prototypes, and user/item embeddings of each layer are updated by aggregating the global user/item/relation/entity signals.

Formally, we could get intent-aware user/item representations at the $l$-th user/item embedding layer, by aggregating information across different $K$ learnable intent prototypes (including $\textbf{c}_{u}$ and $\textbf{c}_{v}$), using the following design: 

\begin{align}\label{intent_integrate}
&\textbf{e}^{l}_{u} = \sum_{k}^{K} \textbf{c}_{u}^{k} P(\textbf{c}_{u}^{k}|\textbf{e}^{l}_{u}),\\
& P(\textbf{c}_{u}^{k}|\textbf{e}^{l}_{u}) = \frac{\eta(\textbf{e}^{l-1\top}_{u}\textbf{c}_{u}^{k})}{\sum_{k'}^{K} \eta(\textbf{e}^{l-1\top}_{u} \textbf{c}_{u}^{k'})},
\end{align}
where the $P(\textbf{c}_{u}^{k}|\textbf{e}^{l}_{u})$ and $P(\textbf{c}_{v}^{k}|\textbf{e}^{l}_{v})$ denotes the importance score of $\textbf{c}_{u}^{k}$ for $l-$th user embeddings that has encodes the global signals. Similarly, the $P(\textbf{c}_{v}^{k}|\textbf{e}^{l}_{v})$ denotes the importance score of $\textbf{c}_{u}^{k}$ for $l-$th item embeddings.

As for the way of calculating the $l-$th user/item embeddings, we propose to adopt a two-step process to encode the global user/item/ relation/entity information in the whole heterogeneous graph.
The first step is to merge the knowledge information (including both relation and entity) into item embeddings with a proposed relation-aware graph aggregation, making the item representation more comprehensive and informative. It injects the relational context into the embeddings of the neighboring entities, and weighting them with the knowledge rationale scores  (It’s worth noting that items are a subset of knowledge entities), as follows:
\begin{equation}\label{beta_cal}
\begin{aligned}
    \vec{e}_{i}^{(l+1)}&=\frac{1}{\left|\mathcal{N}_{i}\right|} \sum\limits_{(r, v) \in \mathcal{N}_{i}} \beta(i, r, v) \vec{e}_{r} \odot \vec{e}_{v}^{(l)},\\
\beta(i, r, v) &=\operatorname{softmax}\left(\left(\vec{e}_{i} || \vec{e}_{r}\right)^{T} \cdot \left(\vec{e}_{v} || \vec{e}_{r}\right) \right) \\
&=\frac{\exp \left(\left(\vec{e}_{i} || \vec{e}_{r}\right)^{T} \cdot \left(\vec{e}_{v} || \vec{e}_{r}\right)\right)}{\sum\limits_{\left(v^{\prime}, r\right) \in \hat{N}(i)} \exp \left(\left(\vec{e}_{i}|| \vec{e}_{r}\right)^{T} \cdot \left(\vec{e}_{v^{\prime}} || \vec{e}_{r}\right)\right)},
\end{aligned}
\end{equation}
where $||$ denotes concat operation, $N_i$ denotes the set of neighboring entities.

Then the second step is to apply a novel graph transformer among user-item graph, which encodes global user/item/entity information into user/item representations. By doing so, the user/item representations of each layer are integrated with global signals, which would be exploited into intent modeling and representation updating, as follows:

\begin{align}
    \label{eq:graphTrans}
    \vspace{-0.4cm}
    &\textbf{e}_u^{l+1} = \sum_{i} \mathop{\Bigm|\Bigm|}\limits_{h=1}^H m_{u,i} \beta_{u, i}^h  \textbf{W}_\text{V}^h \textbf{e}_{i}^l;~~~ m_{u,i}=\left\{
    \begin{aligned}
    &1~~\text{if}~(u,i)\in{\mathbf{Y}}\\
    &0~~\text{otherwise}
    \end{aligned}\right.
    \nonumber\\
    & \beta^h_{u,i} = \frac{\exp \bar{\beta}^h_{u,i}}{\sum_{i}\exp\bar{\beta}^h_{u,i} };~~~~~~~~
    \bar{\beta}^h_{v,v'} = \frac{(\textbf{W}_\text{Q}^h \cdot \textbf{e}_u^{l})^\top \cdot (\textbf{W}_\text{K}^h \cdot \textbf{e}_{i}^{l})}{\sqrt{d/H}},
\end{align}
where $H$ denotes the number of attention heads (indexed by $h$). $m_{v,v'}$ is the binary indicator to decide whether to calculate the attentive relations between user $u$ and item $i$. $\beta_{u,i}^h$ denotes the attention weight for user-item interaction pair $(u,i)$ \wrt\ the $h$-th head representation space. $\textbf{W}_\text{Q}^h, \textbf{W}_\text{K}^h, \textbf{W}_\text{V}^h \in \mathbb{R}^{d/H\times d}$ denotes the query, key, the value embedding projection for the $h$-th head, respectively.

By integrating global information into users/items, we could learn intent-aware user/item representations and update the learnable intents according to Equation~\ref{intent_integrate}.





\subsection{Knowledge Contrastive Denoising under Intents}

It is intuitive that noisy or irrelevant connections between entities in knowledge graphs can lead to suboptimal representation learning, which is opposite to original purpose of introducing the KG. To eliminate the noise effect in the KG and distill informative signals that benefit the recommendation task, we propose to highlight important connections consistent to user intents, while removing the irrelevant ones.

\subsubsection{Knowledge Sampling under intents.}
With the intent-aware user/item representations, we then try to denoise the item-entity graph by removing the irrelevant edges and nodes and sampling the important ones. We first exploit the intent-aware representations to calculate the importance score of knowledge triplets (\ie the item-relation-entity pairs) same as Equation \ref{beta_cal}, then add the Gumbel noise \cite{jang2017categorical} to the learned importance scores to improve the sampling robustness, as follows:

\begin{equation}
    \begin{aligned}
\beta(i, r, v) &=\operatorname{softmax}\left(\left(\vec{e}_{i} || \vec{e}_{r}\right)^{T} \cdot \left(\vec{e}_{v} || \vec{e}_{r}\right) \right) \\
\beta(i, r, v)  &= \beta(i, r, v) -\log\left(-\log(\epsilon)\right);\quad\epsilon\sim\text{Uniform}\left(0,1\right),
    \end{aligned}
\end{equation}
where $\epsilon$ is a random variable sampled from a uniform distribution.
Then it follows a top-k sampling strategy for generating the new item-entity graph that removes the irrelevant edges and nodes:
\begin{equation}
    \widehat\beta(i, r, v) =\left\{\begin{array}{ll}
\beta(i, r, v), & \beta(i, r, v) \in \text { top-k} \left(\beta(i, r, v)\right), \\
0, & \text {otherwise},
\end{array}\right.
\end{equation}
where $\widehat\beta(i, r, v)$ is the sampled triples in item-entity graph, which would be used to replace the original graph structure in the following user/item representation learning.
 

\subsubsection{Local-Global Knowledge Contrastive Learning}

With the sampled item-entity graph, we then propose to iteratively update the intent-aware representations in it. And inspired by previous contrastive learning based methods that align the item representations from KG and CF to denoise, we further propose a local-global contrastive mechanism to improve the robustness of representation learning.

Specifically, we exploit the user-item graph and sampled item-entity graph to perform light information aggregation with intent-aware user/item representations $\textbf{e}_u, \textbf{e}_i$ as input $\textbf{z}_u^{(0)}, \textbf{z}_i^{(0)}$, for acquiring a robust and effective intent-aware user/item representations, as follows:

\begin{equation}
    \begin{array}{l}
\vec{z}_{i}^{(l+1)}=\frac{1}{\left|\mathcal{N}_{i}\right|} \sum\limits_{(r, v) \in \mathcal{N}_{i}}  \vec{e}_{r} \odot \vec{z}_{v}^{(l)}, \\
\vec{z}_{u}^{(l+1)}=\frac{1}{\left|\mathcal{N}_{u}\right|} \sum\limits_{i \in \mathcal{N}_{u}} \vec{z}_{i}^{(l)},
\end{array}
\end{equation}
where $\textbf{z}_u^{(0)}, \textbf{z}_i^{(0)}$ memorize the global signals, and we hence get final representations of user/item $ \vec{z}_{u}^{(l)}, \vec{z}_{i}^{(l)} (l \in L) $.

Besides the supervised user/item representation learning, we propose to perform a contrastive learning between the nodes embeddings that encode global signals and local signals, which is different from traditional cl-based methods that contrast the CF and KG parts.
We perform information aggregation in the sampled graph with the initial user/item representations $\textbf{e}_u^{(0)}, \textbf{e}_i^{(0)}$ to acquire the local results  $ \vec{z}_{u, local}^{(l)}, \vec{z}_{i, local}^{(l)} (l \in L) $, while utilizing the intent-aware user/item representations $\textbf{e}_u, \textbf{e}_i$ that contains global signals to acquire the global results $ \vec{z}_{u}^{(l)}, \vec{z}_{i}^{(l)} (l \in L) $. Then perform layer-wise contrastive learning between local and global results.




The local aggregation layer embeddings $\vec{z}_{u, local}^{(l)}, \vec{z}_{i, local}^{(l)}$ and global aggregation layer embeddings $\vec{z}_{u}^{(l)}, \vec{z}_{i}^{(l)}$ are made to be contrasted in a layer-wise way. We generate each positive pair using the embeddings of the same user (item) from the local view and each of the global view, and other nodes form the negative pairs. 
We could get the contrastive loss of users as follows:
\begin{equation}
\begin{array}{ll}
    \mathcal{L}_{c}^u= \frac{1}{L} \sum\limits_{l=0}^L
    -\log \frac{exp({\operatorname{s}\left({{\vec{z}_{u}^{l}}}, { {\vec{z}_{u,local}^{l}}}\right) / \tau })}
    { \sum\limits_{k \neq u}exp({\operatorname{s}\left({{\vec{z}_{u}^{l}}}, {{\vec{z}_{k}^{l}}} \right)/ \tau }) +
    \sum\limits_{k \neq u} exp({\operatorname{s}\left( {{\vec{z}_{u}^{l}}}, { {\vec{z}_{k, local}^{l}}}\right) / \tau}) },
    \end{array}
\label{loss}
\end{equation}
where $\operatorname{s}(\cdot)$ denotes the cosine similarity calculating, and $\tau$ denotes a temperature parameter. And similarly we could get the contrastive loss of item $\mathcal{L}_{c}^i$. By summing the two contrastive losses we hence have the total local-global contrastive loss $\mathcal{L}_{c}$.

\subsection{Model Prediction}
After learning intent-aware user/item representations with global signals and performing contrastive learning between local and global information, we have multi-layer intent-aware representations for user/item. By summing all the layers' representations, we have the final user/item representations and predict their matching score through inner product, as follows:
\begin{equation}
\begin{array}{l}
    \vec{z}_u = \vec{z}_{u}^{(0)}+\cdots+\vec{z}_{u}^{(K)}, \quad \vec{z}_i=\vec{z}_{i}^{(0)}+\cdots+\vec{z}_{i}^{(K)}.\\
    \hat{y}(u, i)=\vec{z}_{u}^{\top} \vec{z}_{i}.
\end{array}
\end{equation}

By adopting a BPR loss \cite{rendle2012bpr} to reconstruct the historical data, which encourages the prediction scores of a user’s historical items to be higher than the unobserved items, we acquire the supervised loss:

\begin{equation}
    \mathcal{L}_{\mathrm{BPR}}=\sum_{(u, i, j) \in O}-\ln \sigma\left(\hat{y}_{u i}-\hat{y}_{u j}\right),
\end{equation}
where $\boldsymbol{O}=\left\{(u, i, j) \mid(u, i) \in \boldsymbol{O}^{+},(u, j) \in \boldsymbol{O}^{-}\right\}$ is the training dataset consisting of the observed interactions $\boldsymbol{O}^{+}$ and unobserved counterparts $\boldsymbol{O}^{-}$; $\sigma$ is the sigmoid function. 

\subsection{Multi-task Training}

To combine the recommendation task with the self-supervised task, we optimize the whole model with a multi-task training strategy. We combine the local-global contrastive loss with BPR loss, and learn the model parameter via minimizing the following objective function:
\begin{equation}
     \mathcal{L}_{KGTN} = \mathcal{L}_{\mathrm{BPR}} +  \alpha\mathcal{L}_{c}  + \lambda\|\Theta\|_{2}^{2},
\end{equation}
where $\Theta$ is the model parameter set, $\alpha$ is a hyperparameter to determine the local-global contrastive loss ratio, $\beta$ and $\lambda$ are two hyperparameters to control the contrastive loss and $L_2$ regularization term, respectively.




\section{Experiment}

\begin{table}[t]
\centering
\normalsize
\setlength{\tabcolsep}{0.1mm}{
\begin{tabular}{cl|ccc}
\hline
\multicolumn{1}{l}{}                                                                & & \multicolumn{1}{l}{Book-Crossing} & \multicolumn{1}{l}{MovieLens-1M} & \multicolumn{1}{l}{Last.FM} \\ \hline \hline
\multirow{3}{*}{\begin{tabular}[c]{@{}c@{}}User-item   \\ Interaction\end{tabular}} 
& \# users  & 17,860 & 6,036 & 1,872     \\
& \# items & 14,967 & 2,445 & 3,846                       \\
& \# interactions & 139,746 & 753,772 & 42,346                      \\ \hline
\multirow{3}{*}{\begin{tabular}[c]{@{}c@{}}Knowledge\\ Graph\end{tabular}} & \# entities & 77,903 & 182,011 & 9,366                       \\
& \# relations & 25 & 12 & 60                          \\
& \# triplets & 151,500 & 1,241,996 & 15,518     \\ \hline
\hline
\end{tabular}}
\caption{Statistics for the three datasets.}
\label{datasets}
\end{table}

\begin{table*}[tb]
\centering
\setlength{\tabcolsep}{4pt}
\begin{tabular}{c|ll|ll|ll}
\hline
\multirow{2}{*}{Model} & \multicolumn{2}{c}{Book-Crossing} & \multicolumn{2}{c}{MovieLens-1M} & \multicolumn{2}{c}{Last.FM} \\
& \multicolumn{1}{c}{\textit{AUC}} & \multicolumn{1}{c}{\textit{F1}} & \multicolumn{1}{c}{\textit{AUC}} & \multicolumn{1}{c}{\textit{F1}} & \multicolumn{1}{c}{\textit{AUC}} & \multicolumn{1}{c}{\textit{F1}} \\
\hline\hline
BPRMF & 0.6583$(-13.18\%)$ & 0.6117$(-7.59\%)$ & 0.8920$(-4.52\%)$ & 0.7921$(-7.21\%)$ & 0.7563$(-13.41\%)$ & 0.7010$(-9.95\%)$    \\
\hline
CKE   & 0.6759$(-11.42\%)$ & 0.6235$(-6.41\%)$ & 0.9065$(-3.07\%)$ & 0.8024$(-6.18\%)$ & 0.7471$(-14.33\%)$ & 0.6740$(-12.65\%)$  \\
RippleNet   & 0.7211$(-6.90\% )$ & 0.6472$(-4.04\% )$ & 0.9190$(-1.82\% )$ & 0.8422$( -2.20\% )$ & 0.7762$(-11.42\% )$ & 0.7025$(-9.80\% )$ \\
\hline
PER   & 0.6048$(-18.53\%)$ & 0.5726$(-11.50\%)$ & 0.7124$(-22.48\%)$ & 0.6670$(-19.72\%)$ & 0.6414$(-24.90\%)$ & 0.6033$(-19.72\%)$          \\
\hline
KGCN     & 0.6841$(-10.60\%)$ & 0.6313$(-5.63\%)$  & 0.9090$(-2.82\%)$ & 0.8366$(-2.76\%)$ & 0.8027$(-8.77\%)$ & 0.7086$(-9.19\%)$          \\
KGNN-LS    & 0.6762$(-11.39\%)$ & 0.6314$(-5.62\%)$ & 0.9140$(-2.32\%)$ & 0.8410$(-2.32\%)$ & 0.8052$(-8.52\%)$ & 0.7224$(-7.81\%)$          \\
KGAT    & 0.7314$(-5.87\%)$ & 0.6544$(-3.32\%)$ & 0.9140$(-2.32\%)$ & 0.8440$(-2.02\%)$ & 0.8293$(-6.11\%)$ & 0.7424$(-5.81\%)$          \\
CKAN    & 0.7420$(-4.81\%)$ & 0.6671$(-2.05\%)$ & 0.9082$(-2.90\%)$ & 0.8410$(-2.32\%)$ & 0.8418$(-4.86\%)$ & 0.7592$(-4.13\%)$      \\
KGIN    & 0.7273$(-6.28\%)$   & 0.6614$(-2.62\%)$ & 0.9190$(-1.82\%)$ &0.8441$(-2.01\%)$ & 0.8486$(-4.18\%)$ & 0.7602$(-4.03\%)$ \\
CG-KGR    & 0.7498$(-4.03\%)$ & 0.6689$(-1.87\%)$ & 0.9110$(-2.62\%)$ & 0.8359$(-2.83\%)$ & 0.8336$(-5.68\%)$ & 0.7433$(-5.72\%)$      \\
 \hline
KGCL  & 0.7453$(-4.48\%)$ & 0.6679$(-1.97\%)$ & 0.9184$(-1.88\%)$ &0.8437$(-2.05\%)$ &0.8455$(-4.49\%)$ &0.7596$(-4.00\%)$     \\
MCCLK & \underline{0.7625}$(-2.76\%)$ &\underline{0.6777}$(-0.99\%)$ &\underline{0.9252}$(-1.20\%)$ &\underline{0.8559}$(-0.83\%)$ &\underline{0.8663}$(-2.41\%)$ &\underline{0.7753}$(-2.43\%)$ \\
 \hline
\textbf{KGTN}    & \textbf{0.7901}* & \textbf{0.6876}* & \textbf{0.9372}* & \textbf{0.8642}* & \textbf{0.8904}* & \textbf{0.7996}*           \\ \hline
\end{tabular}
\caption{The result of $AUC$ and $F1$ in CTR prediction. The best results are in boldface and the second best results are underlined. * denotes statistically significant improvement by unpaired two-sample $t$-test with $p < 0.001$.}
\label{tab:compare}
\end{table*}

\begin{table}[tb]
\centering
\setlength{\tabcolsep}{2pt}
\begin{tabular}{c|ll|ll|ll}
\hline
\multirow{2}{*}{Model} & \multicolumn{2}{c}{Book-Crossing} & \multicolumn{2}{c}{MovieLens-1M} & \multicolumn{2}{c}{Last.FM} \\
& \multicolumn{1}{c}{\textit{R@10}} & \multicolumn{1}{c}{\textit{R@20}} & \multicolumn{1}{c}{\textit{R@10}} & \multicolumn{1}{c}{\textit{R@20}} & \multicolumn{1}{c}{\textit{R@10}} & \multicolumn{1}{c}{\textit{R@20}} \\
\hline\hline
BPRMF & 0.0334 &0.0525 & 0.0939 & 0.1512 & 0.0923 &0.1740  \\
\hline
CKE  &0.0421 &0.0562 & 0.0867 & 0.1364 &0.0780	&0.1532\\
RippleNet &0.0507 &0.0622 &0.1082 &0.1766 &0.0942  &0.1520\\
\hline
PER  &0.0322 &0.0481 &0.0523 &0.1204 &0.0540  &0.1167 \\
\hline
KGCN  &0.0496 &0.0540 & 0.0965 &0.1720 &0.1416 &0.1776 \\
KGNN-LS &0.0422 &0.0526 &0.1286 &0.1757 &0.1312 &0.1933  \\
KGAT &0.0522 &0.0670 &0.1468 &0.2296 &0.1640 &0.2313  \\
CKAN &0.0462 &0.0566 &0.1511 &0.2400  &0.1412 &0.2465       \\
KGIN  &0.0555 &0.0699 &0.1511 &0.2404 &0.1758 &0.2487 \\
CG-KGR  &0.0612	&0.0781 &0.1621	&0.2495 &0.1578	&0.2106\\
 \hline
KGCL  & 0.0679 & 0.0845 & 0.1633 &0.2499 &0.1759 &0.2471  \\
MCCLK & \underline{0.0769}	& \underline{0.0936}  & \underline{0.1642} & \underline{0.2503} & \underline{0.1835} & \underline{0.2598} \\
 \hline

\textbf{KGTN}    & \textbf{0.1060}* & \textbf{0.1275}* & \textbf{0.1841}* & \textbf{0.2826}*  & \textbf{0.2104}* & \textbf{0.3106}*           \\ \hline
\end{tabular}
\caption{The result of $Recall@10$ and $Recall@20$ in top-$K$ recommendation.}
\label{tab:topkcompare}
\vspace{-0.4cm}
\end{table}

Aiming to answer the following research questions, we conduct both offline experiments and online A/B tests on three public datasets and Alibaba online platform:
\begin{itemize}
    \item \textbf{RQ1:} How does KGTN perform, compared to present models?
    \item \textbf{RQ2:} How do the main components in KGTN affect its effectiveness?
    \item \textbf{RQ3:} How do different hyper-parameter settings affect KGTN?
    \item \textbf{RQ4:} How does KGTN perform with noisy injection? 
    \item \textbf{RQ5:} How does KGTN perform in a live system serving billions of users?
\end{itemize}

\subsection{Experiment Settings}
\subsubsection{Dataset and Metrics}

Three benchmark datasets are utilized to evaluate the effectiveness of KGTN: Last.FM \footnote{\url{https://grouplens.org/datasets/hetrec-2011/}}, Book-Crossing \footnote{\url{http://www2.informatik.uni-freiburg.de/~cziegler/BX/}}, and MovieLens-1M \footnote{\url{https://grouplens.org/datasets/movielens/1m/}}. The detailed statistics of them are summarized in Table~\ref{datasets}, which vary in size and sparsity and make our experiments more convincing.
As for the data pre-process, we first follow RippleNet \cite{wang2018ripplenet} to transform their explicit feedback into implicit one, and randomly sample negative samples from his unwatched items with the size equal to his positive ones to construct the negative parts.
As for the sub-KG construction, we follow RippleNet \cite{wang2018ripplenet} and use Microsoft Satori\footnote{\url{https://searchengineland.com/library/bing/bing-satori}} to construct it for MovieLens-1M, Book-Crossing, and Last.FM datasets. Each sub knowledge graph that follows the triple format is a subset of the whole KG with a confidence level greater than 0.9.


We evaluate our method in two experimental scenarios: (1) In click-through rate (CTR) prediction, we apply the trained model to predict each interaction in the test set. We adopt two widely used metrics \cite{wang2018ripplenet, wang2019knowledge} $AUC$ and $F1$ to evaluate CTR prediction. (2) In top-$K$ recommendation, we use the trained model to select $K$ items with the highest predicted click probability for each user in the test set, and we choose Recall@$K$ to evaluate the recommended sets.

\subsubsection{Baselines}
To demonstrate the effectiveness of our proposed KGTN, we compare it with four types of KGR methods: CF-based methods (BPRMF \cite{rendle2012bpr}), embedding-based method (CKE \cite{zhang2016collaborative}, RippleNet \cite{wang2018ripplenet}), path-based method (PER \cite{yu2014personalized}), GNN-based methods(KGCN \cite{wang2019knowledge}, KGNN-LS \cite{wang2019knowledge-aware}, KGAT \cite{wang2019kgat}, CKAN \cite{wang2020ckan}, KGIN \cite{wang2021learning}, CG-KGR \cite{chen2022attentive}), CL-based methods (KGCL\cite{yang2022knowledge}, MCCLK \cite{zou2022multi}).

\subsubsection{Parameter Settings}
We implement our KGTN and all baselines in Pytorch and carefully tune the key parameters. For a fair comparison, we fix the embedding size to 64 for all models, and the embedding parameters are initialized with the Xavier method \cite{glorot2010understanding}. We optimize our method with Adam \cite{kingma2014adam} and set the batch size to 2048. A grid search is conducted to confirm the optimal settings, we tune the learning rate $\eta$ among$\{0.0001, 0.0003,0.001,0.003\}$ and $\lambda$ of $L2$ regularization term among $\{10^{-7}, 10^{-6}, 10^{-5}, 10^{-4}, 10^{-3}\}$. Other hyper-parameter settings are provided in Table~\ref{datasets}. The best settings for hyper-parameters in all comparison methods are researched by either empirical study or following the original papers.

\subsection{Performance Comparison (RQ1)}\label{compare_exp}

We report the empirical results of all methods in Table \ref{tab:compare} and Table \ref{tab:topkcompare}. The improvements and statistical significance test are performed between KGTN and the strongest baselines (highlighted with underline). Analyzing such performance comparison, we have the following observations:
\begin{itemize}[leftmargin=*]
 \item \textbf{Our proposed KGTN achieves the best results.}
    KGTN consistently outperforms all baselines across three datasets in terms of all measures, which achieves significant improvements over the strongest baselines \wrt AUC by 2.76\%, 1.20\%, and 2.41\% in Book, Movie, and Music respectively, and demonstrates its effectiveness.
    We attribute such improvements to the following aspects: 
    (1) By modeling user intents with global signals, KGTN is able to learn user/item representations in a more fine-grained and comprehensive manner;
    (2) The knowledge sampling strategy under intents could remove less relevant knowledge information for a robust representation learning;
    (3) The local-global contrastive learning improves the representation learning in a self-supervised manner, via contrasting the local and global information.
    \item \textbf{Incorporating KG not always benefits recommender system.}
    Comparing CKE with BPRMF, leaving KG untapped limits the performance of BPRMF, which shows the effectiveness of KG information. While PER gets a worse performance than BPRMF, which means that only incorporating suitable knowledge could benefit the model. This fact stresses the importance of knowledge sampling and knowledge denoising.
    \item \textbf{GNN has a strong power of graph learning.}
    Most of the GNN-based methods perform better, suggesting the importance of modeling long-range connectivity for graph representation learning. This fact inspires us to go beyond the local aggregation paradigm, and to consider the global signals.
    \item \textbf{Contrastive Learning is effective.}
    The most recently proposed CL-based methods have the best performance, which shows the effectiveness of incorporating a self-supervised task for improving representation learning. It inspires us to design proper contrastive mechanisms to denoise the knowledge and improve the model performance.

\end{itemize}

\subsection{Ablation Studies (RQ2)}

\begin{figure*}[t] 
    \centering  
    \includegraphics[width=\linewidth]{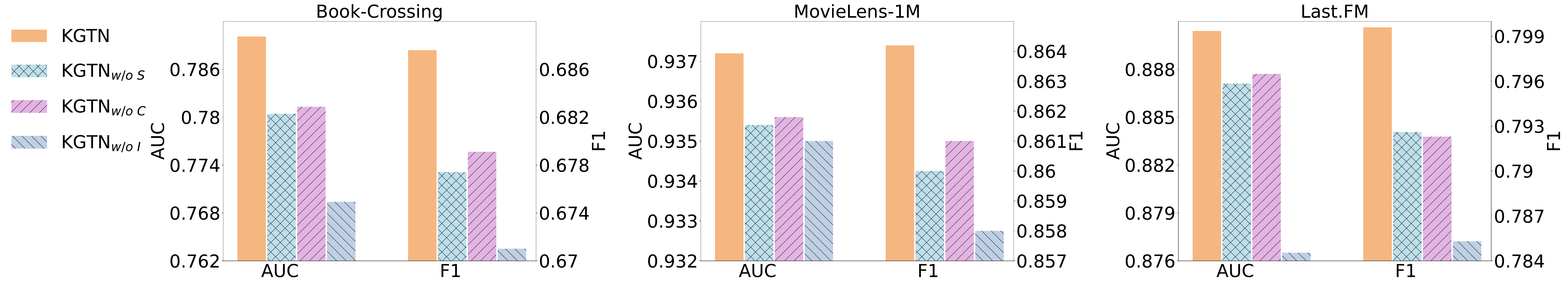}
    \caption{Effect of ablation study.} 
    \label{fig::ablation} 
\end{figure*}
As shown in Figure \ref{fig::ablation}, here we examine the contributions of main components in our model to the final performance by comparing KGTN with the following three variants: 1) $\text{KGTN}_{w/o \ S}$: In this variant, the knowledge sampling under intents module is removed. 2) $\text{KGTN}_{w/o \ C}$: This variant removes local-global contrastive mechanism. 3) $\text{KGTN}_{w/o \ I}$: This variant removes the multi-intent modeling, which means both global intent modeling and knowledge contrastive denoising do not exist in this variant.
The results of two variants and KGTN are reported in Figure \ref{fig::ablation}, from which we have the following observations: 
\begin{itemize}
    \item Removing both knowledge sampling and local-global contrasting would degrade model performance, which shows their effectiveness in representation learning.
    \item Ablating the multi-intent modeling brings the worst performance, which shows the importance of incorporating global signals and considering multiple intents.
    
\end{itemize}

\subsection{Sensitivity Analysis (RQ3)}

\subsubsection{Impact of graph transformer depth.}

\begin{table}[t]
\centering
\setlength{\tabcolsep}{3pt}{
\begin{tabular}{l|c c|c c|c c}
\hline
 & \multicolumn{2}{c|}{Book} & \multicolumn{2}{c|}{Movie} & \multicolumn{2}{c}{Music} \\ 
 & Auc & F1 & Auc & F1 & Auc & F1 \\ \hline\hline

$L$=1 & \textbf{0.7901}	&\textbf{0.6876} & \textbf{0.9372}	&\textbf{0.8642} &\textbf{0.8904}	&\textbf{0.7996} \\ 
$L$=2 & 0.7743	&0.6783 & 0.9349	&0.8623 & 0.8834	&0.8068 \\ 
$L$=3 & 0.7603	&0.6709 & 0.9278	&0.8481 & 0.8785	&0.7951 \\ 
\hline
\end{tabular}}
\caption{Impact of graph transformer depth.}
\label{tab::transdepth}
\vspace{-0.6cm}
\end{table}
To study the influence of graph transformer depth, we vary $L$ in range of \{1, 2, 3\} on book, movie, and music datasets. As shown in Table~\ref{tab::transdepth}, KGTN performs best when $L=1$. It convinces that one iteration is enough for integrating the global signals into user/item representations, which shows its low reliance on model depth.

\subsubsection{Impact of intent number $K$.}
\begin{figure}[t] 
    \centering  
    \subfloat[Book]
    {   \begin{minipage}[t]{0.5\linewidth}
            \centering          
            \includegraphics[width=\textwidth]{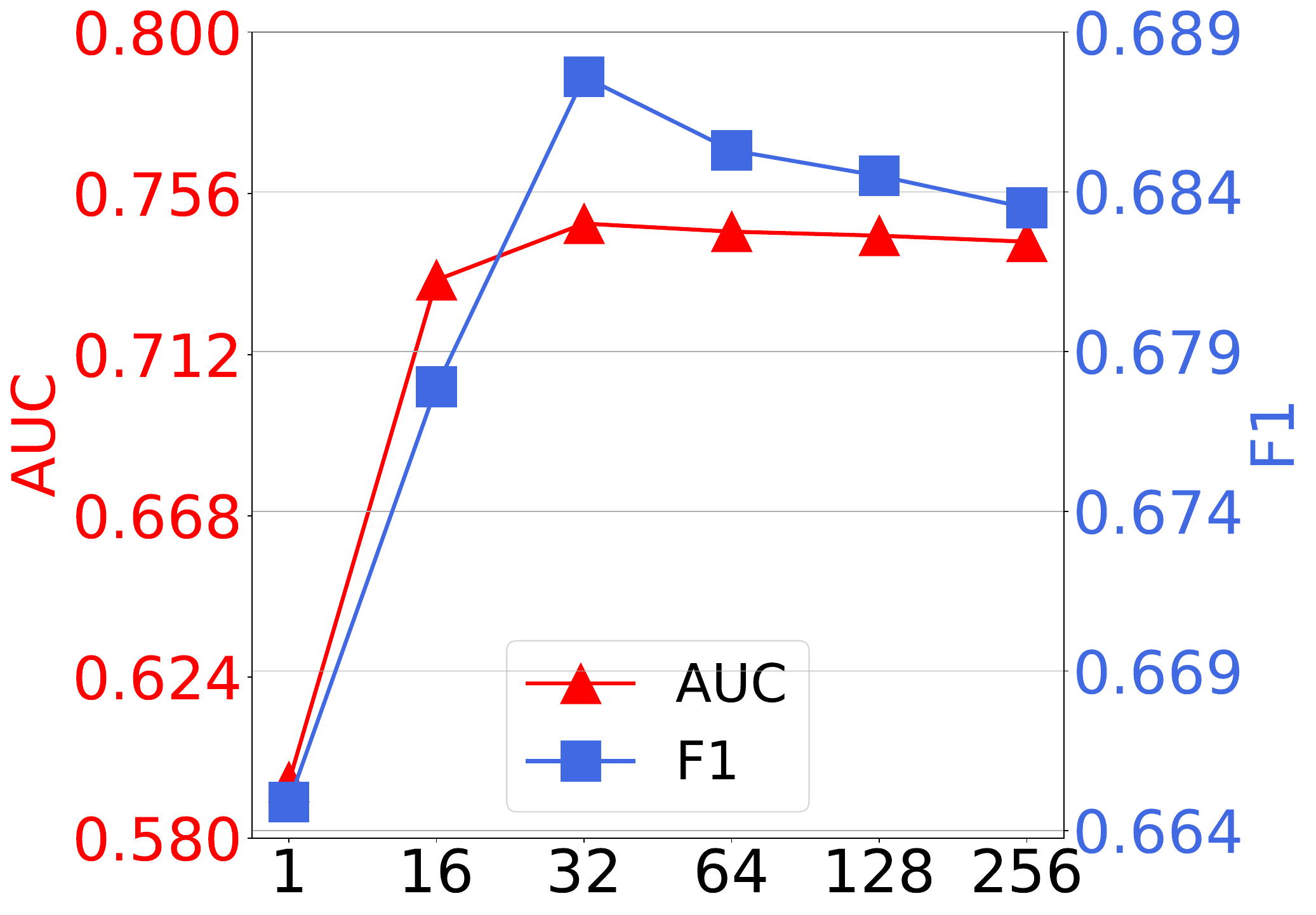} 
        \end{minipage}}
    \subfloat[Music]
    {   \begin{minipage}[t]{0.5\linewidth}
            \centering     
            \includegraphics[width=\textwidth]{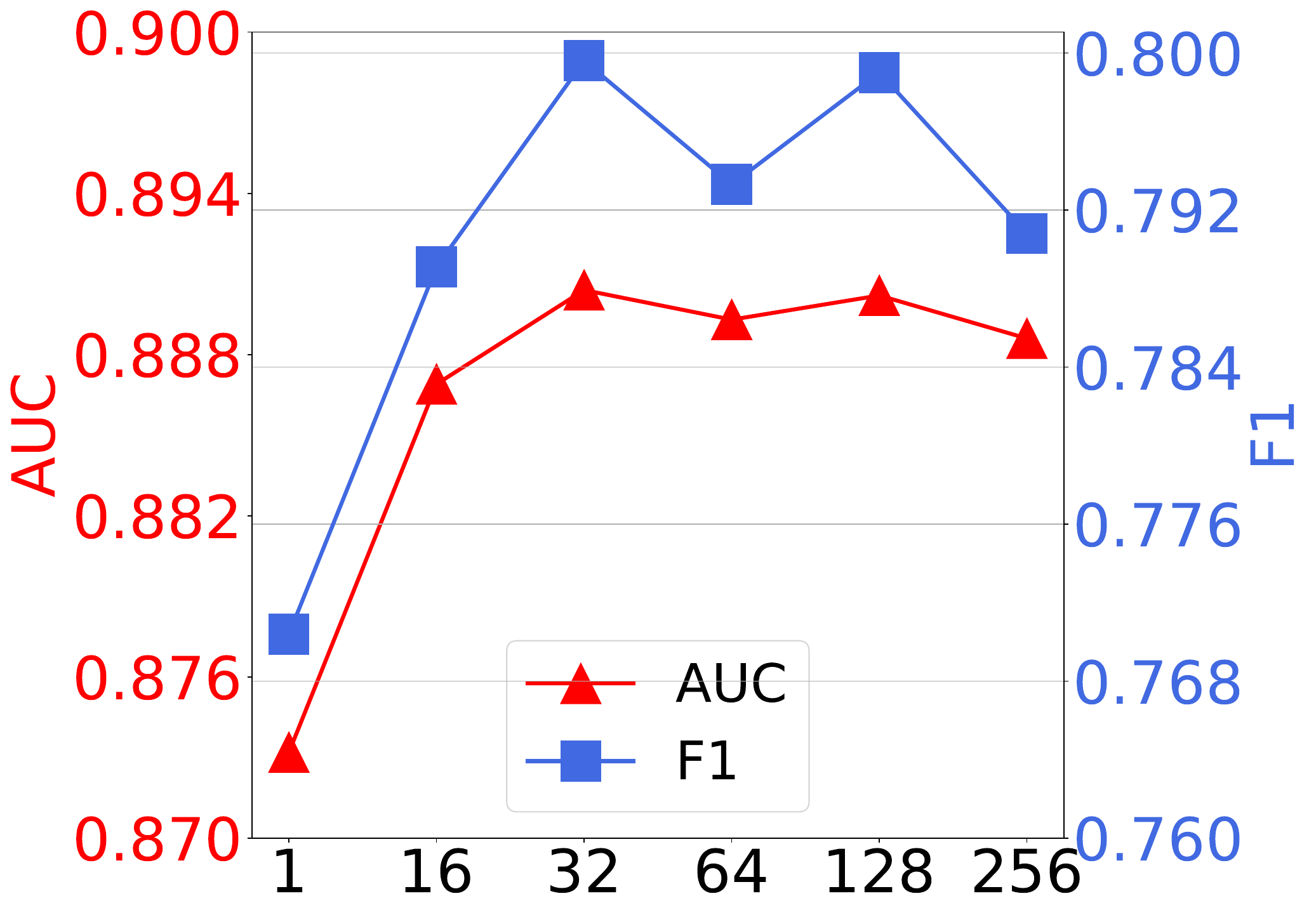} 
        \end{minipage}}
    \vspace{-0.4cm}
    \caption{Impact of intent number $K$.} 
    \label{fig::intent} 
    \vspace{-0.4cm}
\end{figure}

To investigate the impact of the intent number, we vary it from the range \{16, 32, 64, 128, 256\} and the model performance is shown in Figure \ref{fig::intent}, from which we could draw the following conclusions:
i) Ignoring the multiple intents ($K=1$) results in the worst performance, which convinces the effectiveness of incorporating multi-intent modeling.
ii) The model performance first arises then drops with the intent increasing. A suitable intents number boosts the model performance with fine-grained preference learning, while too many intents mean too fine-grained modeling and inversely introduce noise into representation learning.

\subsubsection{Impact of contrastive loss ratio $\alpha$.}
\begin{figure}[t] 
    \centering  
    \subfloat[Book]
    {   \begin{minipage}[t]{0.5\linewidth}
            \centering          
            \includegraphics[width=\textwidth]{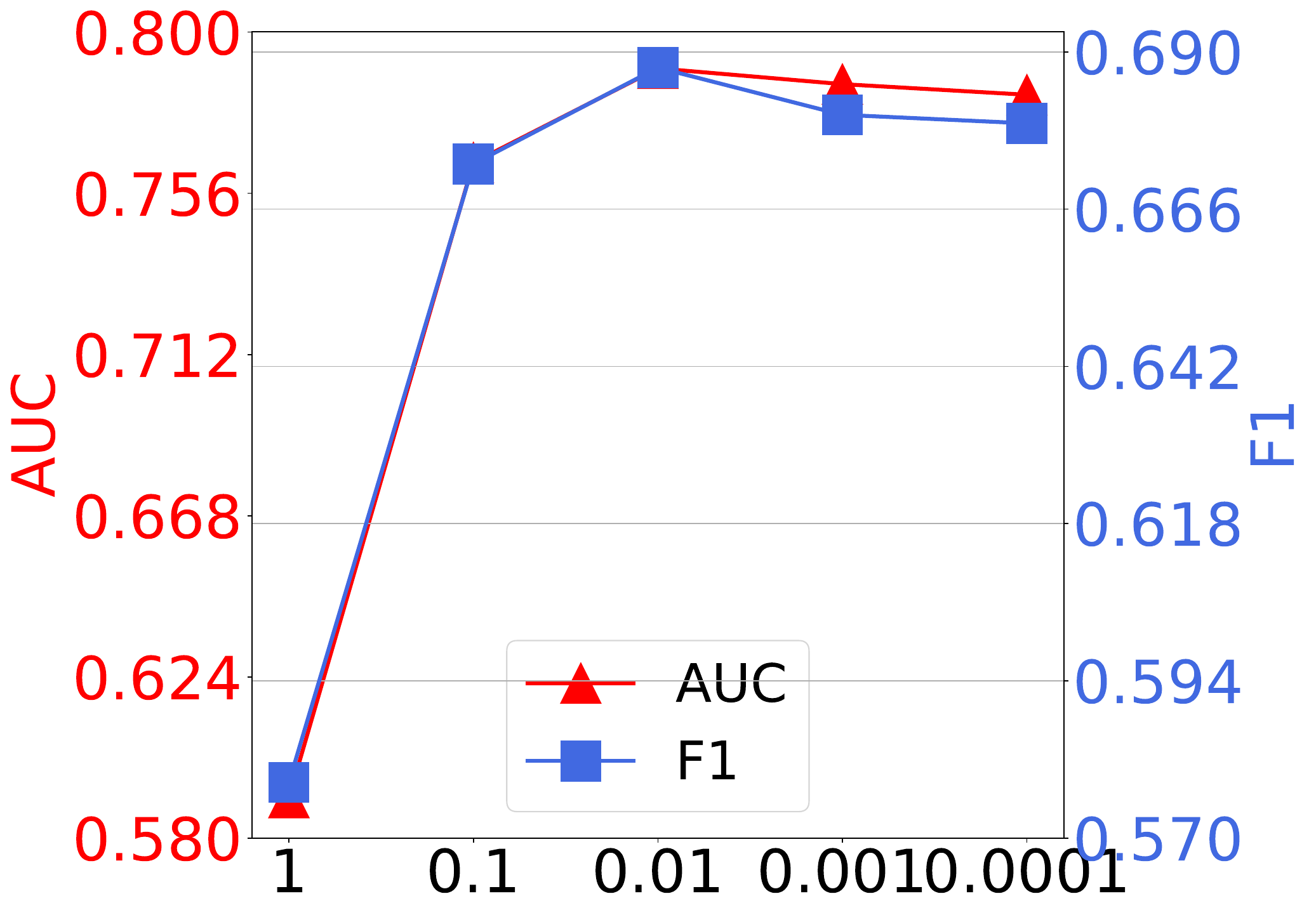} 
        \end{minipage}}
    \subfloat[Music]
    {   \begin{minipage}[t]{0.5\linewidth}
            \centering     
            \includegraphics[width=\textwidth]{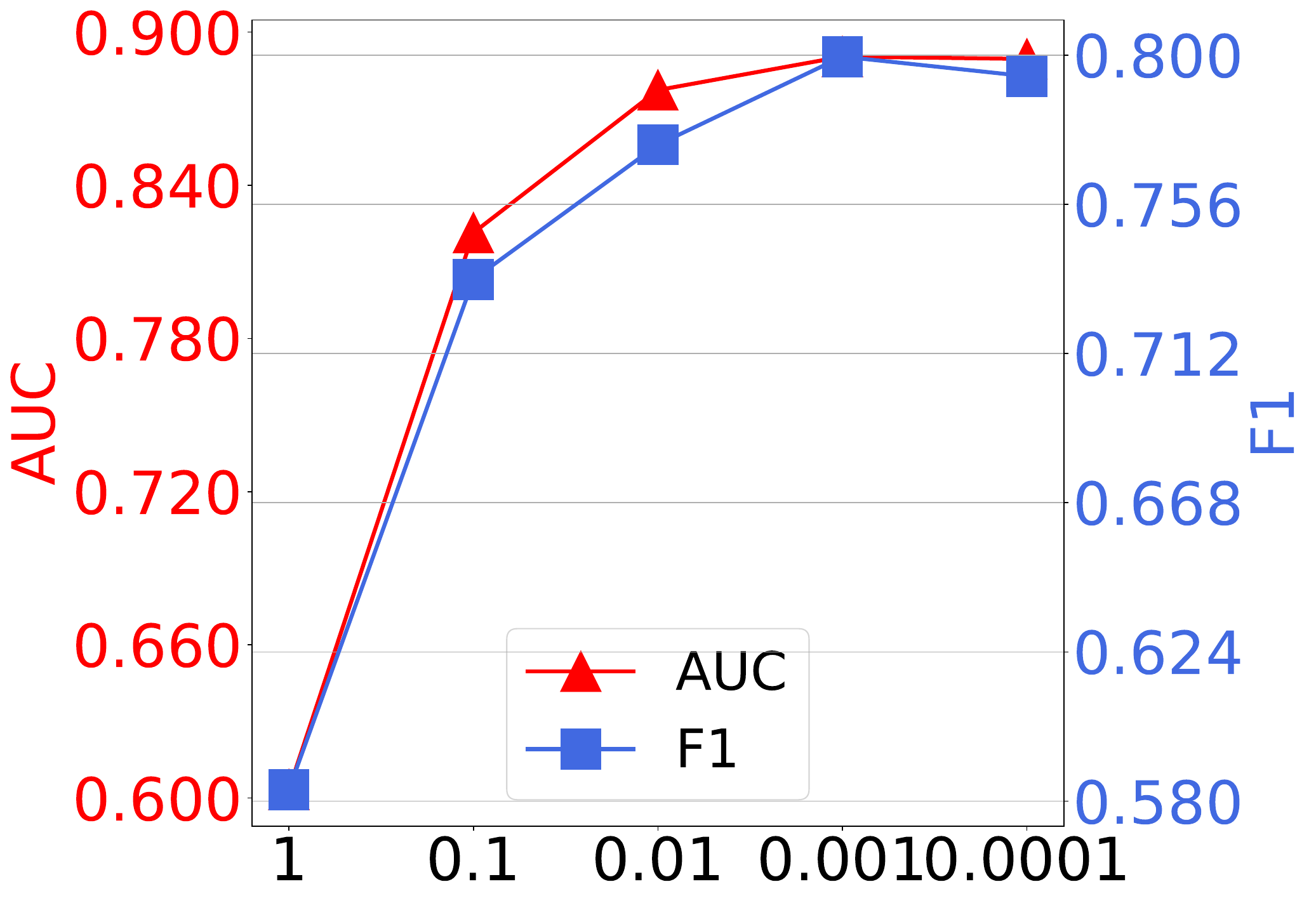} 
        \end{minipage}}
    \vspace{-0.4cm}
    \caption{Impact of contrastive loss ratio $\alpha$.} 
    \label{fig::alpha} 
\end{figure}

The parameter $\alpha$ determines the importance of the contrastive loss during the multi-task training. Hence we vary it in range \{1, 0.1, 0.01, 0.001\} to study its influence. As shown in Figure \ref{fig::alpha}, we observe that: $\alpha=0.1$ brings the best model performance, the main reason is that changing the contrastive loss to a fairly equal level to recommendation task loss could boast the model performance.

\begin{figure}[t] 
    \centering  
    \subfloat[Book]
    {   \begin{minipage}[t]{0.5\linewidth}
            \centering     
            \includegraphics[width=\textwidth]{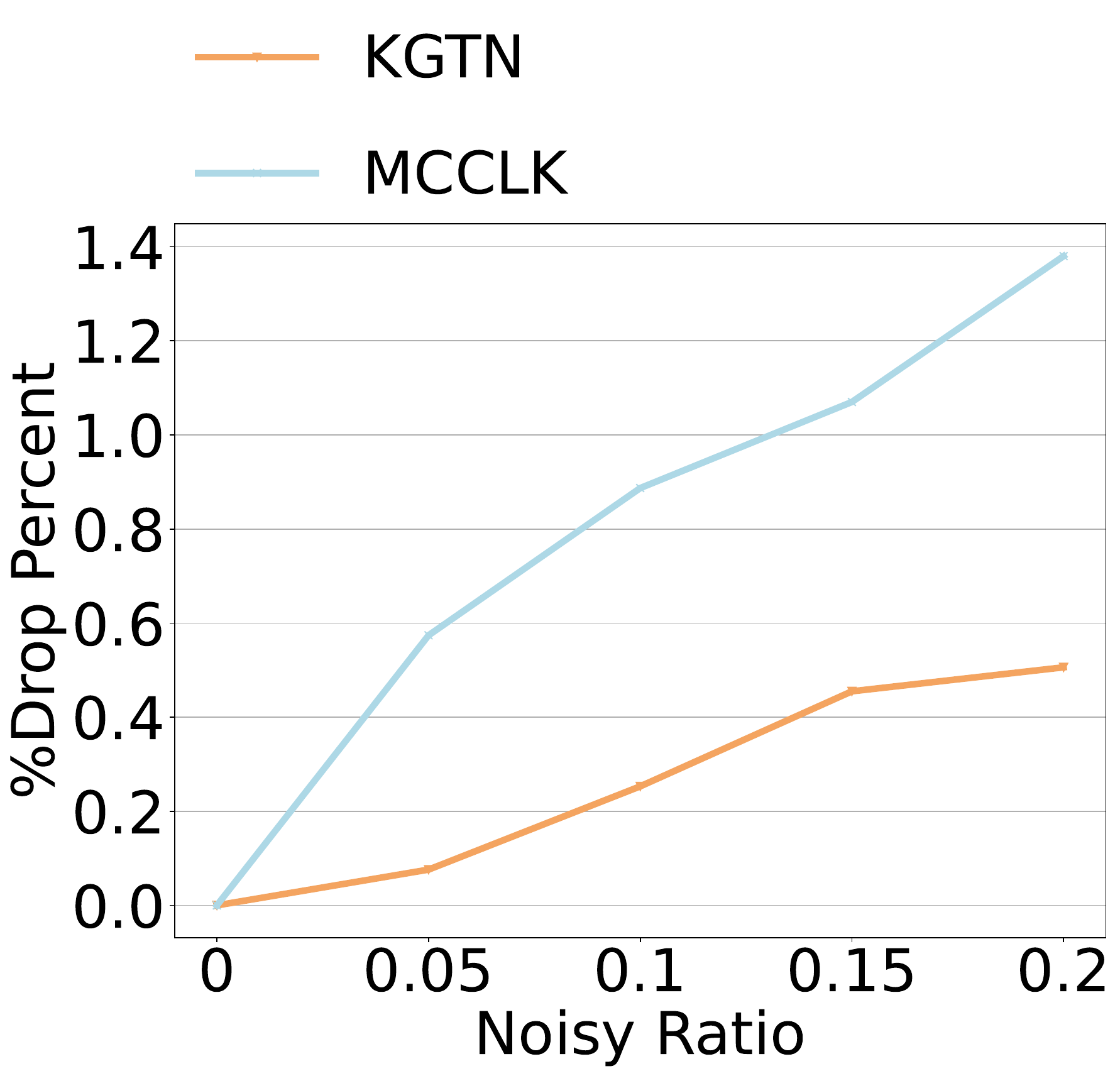} 
        \end{minipage}}
    \subfloat[Music]
    {   \begin{minipage}[t]{0.5\linewidth}
            \centering          
            \includegraphics[width=\textwidth]{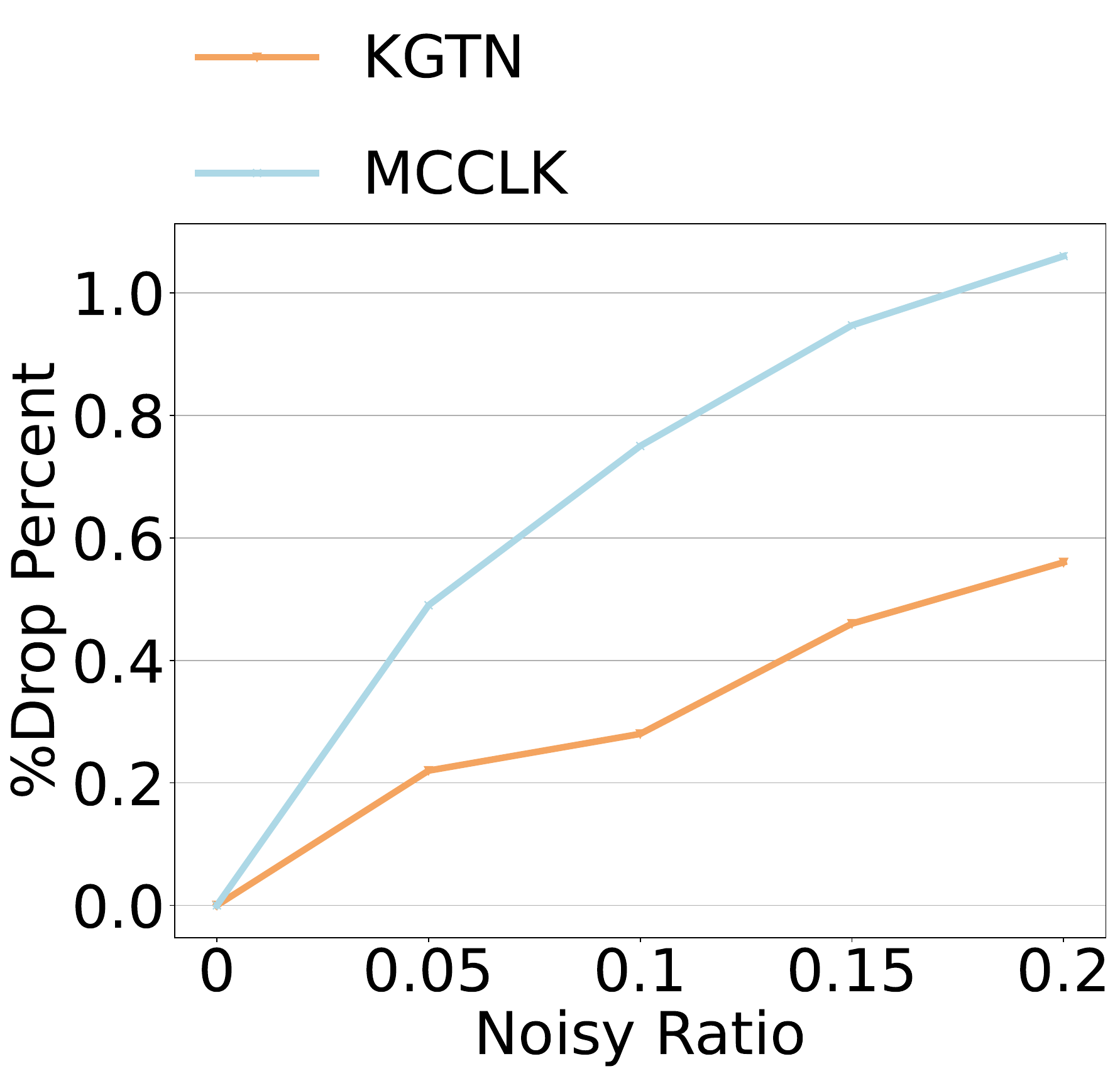} 
        \end{minipage}}
    \caption{Noise Analysis in Music and Book datasets.} 
    \label{fig:noise} 
\end{figure}

\subsection{Denoising Analysis (RQ4)}

We additionally conduct a denoising experiment here, for checking the model robustness under noisy interactions. Specifically, we contaminate the training set by adding a certain proportion of adversarial examples (\ie 5\%, 10\%, 15\%, 20\% negative user-item interactions), meanwhile keeping the validation and testing sets unchanged, following SGL \cite{wu2021self}. This experiment could show the model ability of noise-irrelevant representation learning, from an overall perspective. The experimental results on Baby are shown in Figure~\ref{fig:noise}, where the Noisy Ratio means the percentage of noisy interactions added for the model, and the \%Drop Percentage represents the percentage of performance degradation. 

From the experimental results, we could clearly find that:
Although adding noise degrades the model performance of both KGTN and MCCLK, the proposed KGTN is clearly less influenced than the GNN-based and CL-based MCCLK. It is more apparent with a bigger noise ratio, since the performance dropping gap between KGTN and MCCLK becomes larger and larger as the noise ratio increases, which suggests that KGTN is more robust to noisy perturbation.

\subsection{Online A/B Testing (RQ5)}
We conduct online A/B testing in our recommender system in Alibaba, to validate the benefits of KGTN in the real business scenario. 
In our online A/B testing, users are randomly divided into A group and B group. When group A browses the website, the system recommends results provided by the previous model, while the group B is recommended with results provided by our KGTN model.
Experiments are run for three weeks, during which both the control and experiment models are trained continuously with new interactions and feedback being used as training data. As shown in Table \ref{tab::abtest}, KGTN improves item page view (ipv) per user by 2.3\%, unique visitor list to order (uv l2o) by 2.22\%, and unique visitor click through rate (uv ctr) by 2.3\% relatively compared with DMR \cite{lyu2020deep}, which is the last version of CTR model in our system. This reveals the practical application value of KGTN.

\begin{table}[t]
\centering
\setlength{\tabcolsep}{1pt}{
\begin{tabular}{l|c}
\hline
Metric & Relative Improvement
 \\ \hline\hline

Item page view per user & +2.3\% \\ 
Unique visitor list to order & +2.22\%	   \\ 
Unique visitor click through rate & +2.3\%   \\ 
\hline
\end{tabular}}
\caption{Results of online A/B testing.}
\label{tab::abtest}
\end{table}
\section{Related Work}

\subsection{Knowledge Enhanced Recommendation}
Existing Knowledge Enhanced recommendation methods can be roughly categorized into three lines: embedding-based, path-based, and GNN-based methods.
 \textbf{Embedding-based methods} \cite{wang2018dkn, huang2018improving, zhang2018learning} pre-train the KG entity embeddings with knowledge graph embeddings methods (KGE) \cite{bordes2013translating, lin2015learning}, for enriching item representations, such as CKE \cite{zhang2016collaborative} and KTUP \cite{cao2019unifying}, and RippleNet \cite{wang2018ripplenet}.
\textbf{Path-based methods} \cite{shi2018heterogeneous, zhao2017meta, yu2013collaborative} explore various patterns of connections among items in KG to provide additional guidance for the recommendation, such as PER \cite{yu2014personalized} and MCRec\cite{hu2018leveraging}. KPRN \cite{wang2019explainable} further automatically extracts paths between users and items, modeling these paths with RNNs.
\textbf{GNN-based methods} are founded on the information aggregation mechanism of graph neural networks (GNNs) \cite{hamilton2017inductive, ying2018graph}, which integrates multi-hop neighbors into node representations, modeling long-range connectivity. KGCN \cite{wang2019knowledge} and KGNN-LS \cite{wang2019knowledge-aware} firstly utilize GNN on KG side, then KGAT\cite{wang2019kgat} propose to utilizes GAT on the unified user-item-entity heterogeneous graph. Then CKAN \cite{wang2020ckan} separately models collaborative signals and knowledge signals, and CG-KGR \cite{chen2022attentive} exploits the collaborative signals to guide the aggregation on KG. KGIN \cite{wang2021learning} models user-item interactions at an intent level, which reveals user intents behind the KG interactions and performs GNN on the user-intent-item-entity graph.
More recently, \textbf{CL-based Methods} such as MCCLK \cite{zou2022multi}, KGIC \cite{zou2022improving}, and KGCL \cite{yang2022knowledge} combine a contrastive learning paradigm wtih GNN-based methods, and build cross-view contrastive frameworks as additional self-discrimination supervision signals to enhance robustness.

\subsection{Multi-intent Modeling}
Current multi-intent modeling usually adopts a disentangled representation learning paradigm, which splits the user embedding into multiple chunks and tries to learn each chunked emebedding respectively for representing each user intent. In graph representation learning area, DisenGCN \cite{ma2019disentangled}, IPGDN \cite{liu2020independence} and ADGCN \cite{zheng2021adversarial} adopt such a paradigm and utilize the Hilbert-Schmidt Independence Criterion (HSIC) and adversarial learning for ensuring the effectiveness of intent modeling.
As for recommendation scenarios, DGCF \cite{wang2020disentangled} proposes to disentagnle the user representations and adopts a mutual information restraint for the independence of all intents.
And KGIN \cite{wang2021learning} considers each intent as an attentive combination of KG relations, then use a local aggregation manner for the intent modeling.
MIDGN \cite{zou2023towards} performs fine-grained intent disentanglement from the hierarchical structure in bundle recommendation, together with an intent contrastive mechanism.
Despite the success of disentangled learning attempts in previous methods, they commonly learns the multi-intent representations with local information, while ignore the importance of global signals. Hence, our work focuses on learning intent-aware representations with global information, and exploits the intent features to denoise the knowledge information.

\section{Conclusion}
In this paper, we focus on modeling multiple intents with global information, and leveraging intents to denoise the knowledge information. We propose a novel framework, KGTN, which achieves fine-grained and robust user/item representation learning from two dimensions:
1) KGTN models global intents and learns intent-aware user/item representations with a proposed graph transformer, by integrating global signals into learnable intents.
2) KGTN exploits the user intents to sample the relevant knowledge, and designs a local-global contrastive mechanism within the sampled graph, for robust representation learning.
Extensive experiments on three public datasets demonstrate that KGTN significantly improves the recommendation performance over baselines on both Click-Through rate prediction and Top-K recommendation tasks. Furthermore, the online A/B testing on recommender system of Alibaba demonstrates the practical application value of KGTN.

\section*{Acknowledgments}
This work was supported in part by the National Natural Science Foundation of China under Grant No. 62276110, No. 62172039 and in part by the fund of Joint Laboratory of HUST and Pingan Property \& Casualty Research (HPL). The authors would also like to thank the anonymous reviewers for their comments on improving the quality of this paper. 
\balance

\bibliographystyle{ACM-Reference-Format}
\bibliography{acmart}

\end{document}